\documentclass[10pt,a4paper]{article}
\usepackage{jheppub_kim}
\usepackage{pdflscape}
\usepackage{amsmath}
\usepackage{amssymb}
\usepackage{dcolumn}
\usepackage{bm}
\usepackage{color}
\usepackage{epsfig}
\usepackage{amsfonts}
\usepackage{graphicx}
\usepackage{subfigure}
\usepackage{dcolumn}
\UseRawInputEncoding

\begin{document}

\title{Reconstruction of $f(T, \mathcal{T})$ Lagrangian for various
cosmological scenarios}

\author[a]{Tuhina Ghorui,}
\author[b]{Prabir Rudra,}
\author[c]{Farook Rahaman}

\affiliation[a]{Department of Mathematics, Jadavpur University,
Kolkata-700 032, India.}

\affiliation[b] {Department of Mathematics, Asutosh College,
Kolkata-700 026, India.}

\affiliation[c] {Department of Mathematics, Jadavpur University,
Kolkata-700 032, India.}

\emailAdd{tuhinaghorui.math@gmail.com}
\emailAdd{prudra.math@gmail.com}
\emailAdd{farookrahaman@gmail.com}

%\myclassification{04.20.Dw, 04.20.Ex, 04.20.Cv, 04.70.Bw}

\abstract{In this paper we explore a reconstruction scheme in the
background of the $f(T,\mathcal{T})$ gravity theory for different
cosmological scenarios, where $T$ is the scalar torsion and
$\mathcal{T}$ is the trace of the energy-momentum tensor. Using
the reconstruction technique $f(T, \mathcal{T})$ Lagrangian is
constructed for different cosmological eras such as dust, $\Lambda
CDM$, perfect fluid, etc. Both minimal and non-minimal matter
coupled models are considered for this purpose. Different
cosmological scenarios such as power law expansion, de-Sitter
expansion, etc. have been considered, and using them Lagrangian
functionals are constructed. Mathematical viabilities of all the
constructed functionals have been investigated. The physical
implications of the obtained solutions are discussed in detail. To
check the cosmological compatibility of the constructed
$f(T,\mathcal{T})$ functionals we have generated plots of
important parameters like the equation of state parameter and
deceleration parameter. It is seen that the reconstructed models
are perfectly compatible with the late time accelerated expansion
of the universe.}


\maketitle

\section{Introduction}
In the beginning, it was believed that the universe expanded due
to the momentum gained from the big bang. At present the most
comprehensive theory of gravity accessible to us is the theory of
General Relativity(GR) \cite{gr1} formulated by Albert Eienstein
in 1916. It represents one of the most fundamental discoveries of
modern physics. After a period of extensive research GR has been
more and more easy to handle. With many of its predictions coming
true GR is now the cornerstone of gravity research, astronomy,
astrophysics and cosmology. In the current cosmological scenario,
there are two fundamental questions that have triggered the main
object of theoretical and observational physics. One of them is
symbolized by the dark matter, an important but invisible
component in the current universe that interacts gravitationally
on massive galaxies. The other fundamental question of modern
cosmology is represented by the phenomenon of dark energy. At the
turn of the last century two separate observational studies from
far-off type Ia Supernovae (SNIa) demonstrated that the universe
is expanding at an accelerated rate. With this observation the
concepts of dark energy and dark matter came into existence
\cite{SNIa,snia}. In view of the attractive nature of gravity, the
existence of the dark energy phenomenon signified a paradigm shift
in cosmology and theoretical physics. Since then, numerous
theoretical hypotheses that can explain the dynamics and core
characteristics of this phenomena have been put forth in the
scientific literature \cite{cha}. According to the most recent
astrophysical measurements, the universe is composed of roughly
$27\%$ dark matter, $68\%$ dark energy, and around $5\%$ regular
baryonic matter \cite{bsf,pnc}.

Under the scope of general relativity the late universe is assumed
to be dominated by a cosmological constant. On the other hand the
galaxies and the galaxy clusters ensures the existence of a large
amount of dark matter which keeps a strong gravitational
attraction between the galaxies and prevents them from breaking
apart. These two components together make up the $\Lambda$CDM
model, which can explain both the late time acceleration as well
the early time radiation dominated epoch of the universe. The
$\Lambda$CDM model is the simplest model that can explain the
dynamics of the universe as observed on the basic level. But the
model is not free from pathologies. The $\Lambda$CDM model suffers
from the cosmological constant problem \cite{wein} which is the
disparity between the observed value of the cosmological constant
and the value realized from the quantum regime. The model also
fails to resolve the singularity issues surrounding black holes
and the big bang and also seems to have a growing tension in some
parameters in the late universe \cite{agha}. These issues serve as
motivation to modify the gravitational framework, and give rise to
modified gravity models. The simplest modification is brought
about in the $f(R)$ gravity \cite{frr1, frr2, aa, sv, tt} where
the gravity Lagrangian $R$ is replaced by its arbitrary function
$f(R)$ in the Einstein-Hilbert action. It was seen that apart from
the curvature description of gravity, we can also have a torsion
based equivalent formulation of general relativity. Einstein
developed the teleparallel equivalent of general relativity (TEGR)
\cite{tegr1} where the gravitational field is described by the
torsion tensor and not the usual curvature tensor used in general
relativity. This was an attempt to formulate an unified theory of
gravity and electromagnetism. Technically this is achieved by
using the curvature-less teleparallel connection (in the
Weitzenbock gauge) instead of the torsion-less Levi-Civita
connection. Analogous to the construction of $f(R)$ gravity from
general relativity, one can start from TEGR and construct $f(T)$
gravity by replacing the torsion scalar $T$ by an arbitrary
function $f(T)$ of the torsion scalar. The advantage of modified
teleparallel gravity is that it produces second order field
equations irrespective of the choice of the Lagrangian, and this
fact remains true for a number of extensions of the model as well.
The second order nature means that the gravitational polarization
modes remain identical to those in general relativity \cite{pola}.
As a result of this teleparallel gravity has received a lot of
attention in the recent years in the scientific community and a
lot of work has been done on it \cite{tel1, tel2, tel3, tel4,
tel5, tel6}. Various other investigations in $f(T)$ gravity can be
found in the literature, including cosmological solutions
\cite{aj,prab4,prab5}, late time acceleration \cite{re,kj},
thermodynamics \cite{ij}, cosmological perturbations \cite{sh},
and cosmography \cite{sc}. The reader may refer to \cite{yr} for a
comprehensive review on $f(T)$ gravity.

Since there are no theoretical objections against the coupling of
the geometric part of the action with a non-geometric part, one
could easily proceed with a coupling between the gravitational
sector with standard matter in the action. For curvature based
theories one can easily consider modified theories where the
matter Lagrangian is coupled to functions of the Ricci scalar $R$
\cite{rot1, rot2}. This concept can be extended to arbitrary
functions $f(R,L_m)$, where $L_m$ is the matter lagrangian
\cite{rot3, prab2}. Alternatively, one can consider models where
the Ricci scalar $R$ is coupled with the trace of the energy
momentum tensor $\mathcal{T}$ and extend it to arbitrary
functions, such as in $f(R,\mathcal{T})$ theory \cite{rot4, pp,
prab1, prab3}. We can also consider terms like
$R_{\mu\nu}T^{\mu\nu}$ \cite{rot5}. In the torsion based theories
one can easily perform analogous modifications as discussed above
starting from TEGR and not from GR. In such cases matter can be
coupled to the torsion scalar to obtain theories analogous to the
ones discussed above. One can couple a scalar field non-minimally
with the torsion scalar \cite{rot6}. Coupling the trace of the
energy momentum tensor $\mathcal{T}$ to the torsion scalar $T$ we
can frame the $f(T,\mathcal{T})$ theory \cite{rot7} which is
analogous to $f(R,\mathcal{T})$ theory in the curvature family. We
state that the resulting $f(T,\mathcal{T})$ theory differs from
$f(R,\mathcal{T})$ theory, in that it is a novel modified
gravitational theory, with no curvature-equivalent, and its
cosmological implications prove to be very interesting. We would
also like to mention that these scenarios are different than the
corresponding curvature ones, despite the fact that uncoupled GR
coincides with TEGR. They correspond to novel modified theories,
with a novel cosmological behaviour. These are the reasons why
researchers have looked into this idea in relation to
reconstruction and stability \cite{fc,dr}, late time acceleration
and inflationary phases \cite{tj}, the growth factor of
sub-horizon modes \cite{gj} and quark stars \cite{mj}. Finally we
stress that the above modifications, in which one handles the
gravitational and matter sectors on equal footing, do not present
any problem at the theoretical level, and one would only obtain
observational constraints due to non-geodesic motion.

The cosmological reconstruction approach has been created to
precisely recover the $\Lambda CDM$  properties and know the
expansion history of the universe through updated theories of
gravity. Studies of the physics of such theories are made more
difficult by the complexity of field equations, which makes it
difficult to obtain precise and numerical solution that can be
compared with observations. To determine which class of modified
theory give rise to a specific flat FRW model, one inverts the
field equations using the reconstruction technique, which relies
on the assumption that the expansion history of the universe is
precisely understood. In order to develop a plausible cosmology
that may depict the evolution from the matter-dominated era to DE
phase, the cosmological reconstruction has been carried out in the
framework of $f(R)$ gravity under numerous
scenarios\cite{ss,sa,sd,nj,pk,sr,prab6}. Similarly reconstruction
scheme in the background of $f(Q,\mathcal{T})$ gravity has been
performed in \cite{fqtr}. Motivated by the above discussion here
we are interested in studying a cosmological reconstruction
scenario in the background of $f(T,\mathcal{T})$ gravity theory in
order to understand the background dynamics of the theory in
detail. We speculate that the novel features of the
$f(T,\mathcal{T})$ theory will have deep effect in the
cosmological reconstruction scheme and we will obtain very
interesting results from the study. The paper is organized as
follows. In section 2 we furnish the necessary field equations of
the underlying gravity theory. In section 3 we study the
reconstruction scheme for minimally coupled models. In section 4
we investigate the reconstruction scheme for the non-minimally
coupled models. Some important cosmological solutions are
discussed in section 5. Finally the paper ends with a conclusion
in section 6.

\section{Field Equations of $f(T,\mathcal{T})$ gravity}

As it has been mentioned previously that although uncoupled GR and
TEGR are completely equivalent at all levels, matter coupled forms
of these theories are different at both the background and at the
perturbation level. Matter coupled theories in the torsional
set-up, especially with non-minimal coupling between gravity and
matter is completely non-trivial compared to the matter coupled
theories in the curvature set-up. So $f(T,\mathcal{T})$ theory is
a novel modified gravitational theory, with no
curvature-equivalent, and its cosmological implications prove to
be very interesting. The basic equations for the $f(T)$ and
$f(T,\mathcal{T})$ theories of gravity are presented in this
section. Teleparallel theory is primarily based on the
teleparallel connection, which is given by,
\begin{equation}
\widehat{\Gamma}^{\alpha}_{\mu\nu}=e^{\alpha}_{a}\left(\partial_{\nu}e^{a}_{\mu}+\omega^{a}_{b\nu}e^{b}_{\mu}\right)
\end{equation}
where $e^{a}_{\mu}$ is the tetrad and $\omega^{a}_{b\nu}$ are the
components of the teleparallel spin connection. Here the tetrad is
the fundamental dynamical variable in the teleparallel theory and
the metric tensor line elements are produced by the tetrads. Here
the fundamental field is given by $e_{a}(x^\mu)$, which at each
spacetime point $x^{\mu}$ forms an orthonormal basis for the
tangent space, such that $e_{a}~.~e_{b}=\eta_{ab}$, where
$\eta_{ab}=$~diag$(1, -1, -1, -1)$ is the Minkowski metric.
Moreover in the coordinate basis we can express the vierbiens as
$e_{a}=e^{\mu}_{a}\partial_{\mu}$. These vierbeins relate to the
metric tensor $g_{\mu\nu}$ at each point $x$ of the space-time
manifold as given by,
\begin{equation}\label{flrw}
g_{\mu\nu}(x)=e^{a}_{\mu}(x)e^{b}_{\gamma}(x)\eta_{ab}
\end{equation}

Moreover the modified teleparallel theories employ as their field
variables both the tetrad and the spin connection, meaning that
both of them should be described by respective field equations to
satisfy the local Lorentz invariance. However, these two
variations are closely connected, since the field equations of the
spin connection are the antisymmetric part of the field equations
of the tetrad \cite{spin1, spin2}. It is shown in \cite{spin2}
that the field equations obtained from variations with respect to
the tetrad and the spin connection are related, in the sense that
the second is the antisymmetric part of the first. This is imposed
by the local Lorentz invariance and indicates the fact that the
spin connection is nothing more but a pure gauge degree of
freedom. So it doesn't make any sense in deriving the spin
connection field equations explicitly, as they are redundant
\cite{spin2}.

The gravitational action of teleparallel equivalent of general
relativity is given as,
\begin{equation}\label{flrw5}
S_{TEGR}=\int e[T+\mathcal{L}_{m}]d^{4}x
\end{equation}
Here $e=det(e^{a}_{\mu})=\sqrt{-g}$ and $\mathcal{L}_{m}$ is the
matter lagrangian. As a matter of fact, $T$ can be extended to
$T+f(T)$, which is called the $f(T)$ gravity with the action
$S_{f(T)}=\int e[T+f(T)+\mathcal{L}_{m}]d^{4}x$. It can further be
generalized to include matter coupling to provide the
$f(T,\mathcal{T})$ gravity, which is a function of the torsion
scalar $T$ and the trace of the energy-momentum tensor
$\mathcal{T}$. The gravitational action for $f(T,\mathcal{T})$
gravity is given by,
\begin{equation}\label{flrw6}
S_{f(T,\mathcal{T})}=\frac{1}{16\pi G}\int
e\left[T+f(T,\mathcal{T})\right]d^{4}x+\int e
\mathcal{L}_{m}d^{4}x
\end{equation}
Now we can vary the above action both with respect to the tetrad
as well as the spin connection. Variation of the action with
respect to the tetrad gives the field equation as,
\begin{equation}\label{feld11}
\left(1+f_{T}\right)e\left[e^{-1}\partial_{\mu}\left(eS_{a}^{\lambda
\mu}\right)-T^{\sigma}_{\mu
a}S^{\mu\lambda}_{\sigma}+\omega^{b}_{a\nu}S_{b}^{\nu\lambda}\right]\delta
e_{a}^{\lambda}+ef_{\mathcal{T}}\delta
\mathcal{T}+\left(f+T\right)eE_{\mu}^{a}\delta e_{a}^{\mu}=8\pi G
\theta^{\mu}_{a}
\end{equation}
where $E_{\mu}^{a}$ is the inverse of the tetrad $e_{a}^{\mu}$.
Also $\theta^{\mu}_{a}$ is the energy-momentum tensor of the
matter fields, which occurs from the variation of the matter
Lagrangian. We can also vary the action with respect to the spin
connection, $\omega^{a}_{b\nu}$, but it is seen that the resulting
equation is just an antisymmetric part of the above equation and
so both are related. This is imposed by the local Lorentz
invariance and indicates the fact that the spin connection is
nothing more but a pure gauge degree of freedom.

In particular we make a gauge choice (Weitzenbock gauge) by
considering a vanishing spin connection to get the theory based on
the Weitzenbock gauge connection,\textbf{}
\begin{equation}
\Gamma^{\alpha}_{\mu\nu}=e_{a}^{\alpha}\partial_{\nu}e^{a}_{\mu}
\end{equation}
The Weitzenbock connection leads to zero curvature instead of the
Levi-Civita connection (leading to zero torsion). Therefore the
torsional tensor expanding the gravity field is
\begin{equation}\label{flrw1}
 T^{\alpha}_{\mu\nu}=\Gamma^{\alpha}_{\nu\mu}-\Gamma_{\mu\nu}^{\alpha}=e_{a}^{\alpha}(\partial_{\mu}e^{a}_{\nu}-\partial_{\nu}e^{a}_{\mu}).
\end{equation}
We define respectively the contorsion and the super-potential
tensor by the torsional tensor components,
\begin{equation}\label{flrw2}
K^{\mu\nu}_{\alpha}=-\frac{1}{2}(T^{\mu\nu}_{\alpha}-T_{\alpha}^{\nu\mu}-T_{\alpha}^{\mu\nu})
\end{equation}

\begin{equation}\label{flrw3}
S^{\mu\nu}_{\alpha}=\frac{1}{2}(K^{\mu\nu}_{\alpha}+\delta^{\mu}_{\alpha}T^{\lambda\nu}_{\lambda}-\delta^{\nu}_{\alpha}T^{\lambda\mu}_{\lambda})
\end{equation}
By using equation(\ref{flrw1}) and (\ref{flrw3}), we get the
torsion scalar \cite{yp,tj,jw}.
\begin{equation}\label{flrw4}
T=S_{\alpha}^{\mu\nu}T^{\alpha}_{\mu\nu}=\frac{1}{4}T^{\alpha\mu\nu}T_{\alpha\mu\nu}+\frac{1}{2}T^{\alpha\mu\nu}T_{\nu\mu\alpha}-T_{\alpha\mu}^{\alpha}T^{\nu\mu}_{\nu}
\end{equation}
In the Weitzenbock gauge the field equations (\ref{feld11})
becomes,
$$\left(1+f_{T}\right)\left[e^{-1}\partial_{\mu}\left(ee_{a}^{\alpha}S_{\alpha}^{\lambda\mu}\right)-e_{a}^{\alpha}T^{\mu}_{\nu\alpha}S_{\mu}^{\nu\lambda}\right]+e_{a}^{\lambda}\left(\frac{f+T}{4}\right)
+\left(f_{TT}\partial_{\mu}T+f_{T\mathcal{T}}\partial{\mu}\mathcal{T}\right)e_{a}^{\alpha}S_{\alpha}^{\lambda\mu}$$
\begin{equation}\label{flrw7}
-f_{\mathcal{T}}\left(\frac{e_{a}^{\alpha}T_{\alpha}^{\lambda}
+p_{m}e_{a}^{\lambda}}{2}\right)=4\pi G
e_{a}^{\alpha}T_{\alpha}^{\lambda}
\end{equation}
where $f_{\mathcal{T}}=\frac{\partial f}{\partial \mathcal{T}}$,
$f_{T}=\frac{\partial f}{\partial T}$,
$f_{TT}=\frac{\partial^{2}f}{\partial T^2}$,
$f_{T\mathcal{T}}=\frac{\partial^{2}f}{\partial T
\partial \mathcal{T}}$ and $T_{\alpha}^{\lambda}$ is the torsion tensor. In order to apply the aforementioned theory in a
cosmological setting and produce modified Friedman equations, we
incorporate the flat Friedmann-Lemaitre-Robertson-Walker (FLRW)
metric as usual. The FLRW metric is given by,
\begin{equation}\label{flrw8}
ds^{2}=dt^{2}-a(t)^{2}\delta_{ij}dx^{i}dx^{j}
\end{equation}
where $a(t)$ is the coefficient of scale (cosmological scale
factor). We consider the tetrad compatible with the Weitzenbock
gauge as,
\begin{equation}
e^{a}_{\mu}=diag(1, a(t), a(t), a(t))
\end{equation}

Using the above metric with the field equations (\ref{flrw7}) we
get the modified Friedmann equations for the $f(T, \mathcal{T})$
theory as,
\begin{equation}\label{flrw9}
H^{2}=\frac{8\pi G}{3}\rho_{m}-\frac{1}{6}\left(f+12 H^{2}
f_{T}\right)+f_{\mathcal{T}}\left(\frac{\rho_{m}+p_{m}}{3}\right)
\end{equation}

\begin{equation}\label{flrw10}
\dot{H}=-4\pi G\left(\rho_{m}+p_{m}\right)-\dot{H}(f_{T}-12
H^{2}f_{TT})-H\left(\dot{\rho}_{m}-3\dot{p}_{m}\right)f_{T\mathcal{T}}-f_{\mathcal{T}}\left(\frac{\rho_{m}+p_{m}}{2}\right)
\end{equation}
where $\rho_{m}$ and $p_{m}$ are respectively the energy density
and pressure of matter. We consider matter as perfect fluid
described by the energy-momentum tensor,
\begin{equation}\label{perfectfl}
\mathcal{T}_{\mu\nu}=\left(\rho_{m}+p_{m}\right)u_{\mu}u_{\nu}+p_{m}g_{\mu\nu}
\end{equation}
Moreover $u_{\mu}$ is the four-velocity of the fluid which is
normalized as $u_{\mu}u^{\mu}=-1$. Using the above expression the
trace of energy momentum tensor is obtained as,
\begin{equation}\label{traceemt}
g^{\mu\nu}\mathcal{T}_{\mu\nu}=\mathcal{T}=\rho_{m}-3p_{m}
\end{equation}
The Hubble parameter $H$ is given by $H=\frac{\dot{a}(t)}{a(t)}$.
The torsion scalar for the above metric is obtained as $T=-6H^2$.

Contrasting the modified Friedmann equations (\ref{flrw9}) and
(\ref{flrw10}) with the equations of general relativity we get,
\begin{equation}\label{flrw11}
H^{2}=\frac{8\pi G}{3}\left(\rho_{m}+\rho_{de}\right)=\frac{8\pi
G}{3}\rho_{eff}
\end{equation}

\begin{equation}\label{flrw12}
\dot{H}=-4\pi G\left(\rho_{m}+p_{m}+\rho_{de}+p_{de}\right)=-4\pi
G\left(\rho_{eff}+p_{eff}\right)
\end{equation}
where $\rho_{eff}=\rho_{m}+\rho_{de}$ and $p_{eff}=p_{m}+p_{de}$.
In the above equation the corresponding dark energy components are
given by,
\begin{equation}\label{flrw13}
\rho_{de}=\frac{1}{16\pi
G}\left[-f-12f_{T}H^{2}+2f_{\mathcal{T}}\left(\rho_{m}+p_{m}\right)\right]
\end{equation}
and
\begin{equation}\label{flrw14}
p_{de}=\frac{1}{16\pi
G}\left[f+12f_{T}H^{2}-2f_{\mathcal{T}}\left(\rho_{m}+p_{m}\right)\right]+(\rho_{m}+p_{m})\left[\frac{1+\frac{f_{\mathcal{T}}}{8\pi
G}}{1+f_{T}-12H^{2}f_{TT}+H\left(\frac{d\rho_{m}}{dH}\right)\left(1-3c_{s}^2\right)f_{T\mathcal{T}}}-1\right]
\end{equation}
where $c_{s}^2=\frac{dp_{m}}{d\rho_{m}}$ is the speed of sound
squared. The dark energy equation of state (EoS) parameter is
given by,
\begin{equation}\label{deeos}
\omega_{de}=\frac{p_{de}}{\rho_{de}}
\end{equation}

The effective equation of state parameter is given by
\begin{equation}\label{flrw15a}
\omega_{eff}=\frac{p_{eff}}{\rho_{eff}}=-1-\frac{2\dot{H}}{3H^{2}}
\end{equation}
If we consider a universe filled with dust, then $p_{m}=0$, and
hence the effective EoS is given by
$\omega_{eff}=\frac{\omega_{de}}{1+\left(\rho_{m}/\rho_{de}\right)}$.
Moreover the energy density and pressure components satisfy the
conservation equation,
\begin{equation}
\dot{\rho}_{eff}+3H\left(\rho_{eff}+p_{eff}\right)=0
\end{equation}
which can be expanded as,
\begin{equation}\label{conss}
\dot{\rho}_{de}+\dot{\rho}_{m}+3H\left(\rho_{m}+\rho_{de}+p_{m}+p_{de}\right)=0
\end{equation}
In the above equations it should be noted that the cumulative
components including both the dark energy and matter sectors
satisfy the conservation equation. Thus, one obtains an effective
interaction between the dark energy and matter sectors, which is
usual in modified matter coupling theories \cite{lm, th}. More
importantly the dark energy sector is not conserved alone in this
theory, i.e. $\nabla_{\mu}\mathcal{T}^{\mu}_{\nu}\neq 0$. So there
is an effective coupling between dark energy and normal matter
components, with the possibility of energy transfer from one
component to another. This can be modelled by the introduction of
an interaction term $Q$ in the conservation equation. This
interaction term may depend on various dynamical parameters, the
most common being the energy densities of the components of the
universe. So the dark energy sector alone will satisfy a modified
conservation equation in the following form,
\begin{equation}
\dot{\rho}_{de}+3H\left(\rho_{de}+p_{de}\right)=-Q(\rho_{m},p_{m})
\end{equation}
Here the interaction term Q acts as a source function given by,
\begin{equation}
Q(\rho_{m},p_{m})=\dot{\rho}_{m}+3H\left(\rho_{m}+p_{m}\right)
\end{equation}
So in this model there is an exchange of energy from the ordinary
matter to the dark energy sector, and this can be considered as a
possible mechanism driving the accelerated expansion of the
universe.

The deceleration parameter, which serves as an indicator of the
accelerating universe can be given by,
\begin{equation}\label{decel}
q=-1-\frac{\dot{H}}{H^2}
\end{equation}
It is known that positive values $q$ corresponds to a decelerating
universe, while negative values corresponds to accelerating
evolution. In the following sections we will perform the
cosmological reconstruction scheme using the $f(T,\mathcal{T})$
gravity.

%%%%%%%%%%%%%%%%%%%%%%%%%%%%%%%%%%%%%%%%%%%%%%%%%%%%%%%%%%%%%%
\section{Reconstruction of $f(T,\mathcal{T})$ Lagrangian for minimally matter coupled models}
%%%%%%%%%%%%%%%%%%%%%%%%%%%%%%%%%%%%%%%%%%%%%%%%%%%%%%%%%%%%%

We will now look at the solutions of eqns.(\ref{flrw9}) and
(\ref{flrw10}) from a cosmological stand point. In the following
subsections, we hope to show that any cosmic epoch, whether
dominated by matter, radiation, or dark energy, can be developed
by a model of $f(T,\mathcal{T})$ gravity. Using the model
reconstruction approach, we consider $f(T,\mathcal{T})$ in the
functional form as
\begin{equation}\label{flrw15}
f(T,{\mathcal{T}})=f_{1}(T)+f_{2}(\mathcal{T})
\end{equation}
Where $f_{1}$ is a function of $T$ and $f_{2}$ is a function of
$\mathcal{T}$.

The aforementioned additive separable model incorporates a wide
range of cosmological constrains, including TEGR ($f_{1}=T$  and
 $f_{2}=0$), $\Lambda CDM$ ($f_{1}+f_{2}=2 \Lambda$), $f(T)$
gravity ($f_{2}=0$), and  TEGR with a tweak $(f_{1}\neq{T})$ that
enables the $f_{1}(T)$ and $f_{2}(\mathcal{T})$ functions to
accurately represent the behavior of the effective fluid
component. This kind of model has the benefit of producing a
decoupled system of easier to solve ordinary differential
equations for both the $f_{1}(T)$ and $f_{2}(\mathcal{T})$
functions.

The Friedmann equation (\ref{flrw9}), which produces a separated
partial differential equation as a result of our use of basic
additive model $f(T,\mathcal{T})$ is given by,
\begin{equation}\label{aa}
-T+f_{1}(T)-2Tf_{1T}=16 \pi G
\rho_{m}-f_{2}(\mathcal{T})+2f_{2\mathcal{T}}\left(\rho_{m}+p_{m}\right)=\psi
\end{equation}
Here the subscripts $T$ and $\mathcal{T}$ represent derivatives
with respect to the variables and $\psi$ is a constant. We
consider the generalized case for barotropic fluid satisfying
$p_{m}=\omega\rho_{m}$, where $\omega$ is constant. The trace of
the energy momentum tensor is then given by
$\mathcal{T}=\rho_{m}\left(1-3\omega\right)$.

Using the relation $\rho_{m}=\frac{\mathcal{T}}{1-3\omega}$, the
above equation (\ref{aa}) can be expressed as,
\begin{equation}\label{bb}
-T+f_{1}(T)-2Tf_{1T}=\frac{16 \pi G
\mathcal{T}}{1-3\omega}-f_{2}(\mathcal{T})+\frac{2(1+\omega)}{1-3\omega}
\mathcal{T} f_{2\mathcal{T}}=\psi
\end{equation}
We see that there are independent functions of $T$ and
$\mathcal{T}$, on the left side and right side of the equation
respectively. With the assumption that $\psi\neq0$, we observe
that we get two non-homogeneous differential equations with the
analytical solutions as,
\begin{equation}
f_{1}(T)=-T+\psi+\alpha\sqrt{T}
\end{equation}
and
\begin{equation}
f_{2}(\mathcal{T})=-\psi-\frac{16 \pi G
\mathcal{T}}{1+5\omega}+\beta\left[2\mathcal{T}(1+\omega)\right]^{\frac{1-3\omega}{2\left(1+\omega\right)}}
\end{equation}
where $\alpha$ and $\beta$ are constants of integration.

The solution of $f(T,\mathcal{T})$ model is given as,
\begin{equation}\label{fff}
f(T,\mathcal{T})=-T+\alpha\sqrt{T}-\frac{16 \pi G
\mathcal{T}}{1+5\omega}+\beta\left[2\mathcal{T}(1+\omega)\right]^{\frac{1-3\omega}{2(1+\omega)}}
\end{equation}

We see that the obtained Lagrangian is non-linear in both $T$ and
$\mathcal{T}$ which is an important thing to note. We should also
note that for the torsion sector we have a linear term $T$ and
also a non-linear term $\sqrt{T}$. Same is true for the matter
sector $\mathcal{T}$. Depending on the value of $\omega$ the
matter dependent terms will vary, but the torsion sector will
remain unchanged for the different cosmological epoch. We should
also note that the non-linear terms in $T$ and $\mathcal{T}$
occurs with the constants $\alpha$ and $\beta$. So the effect of
these terms on the Lagrangian can be fine tuned by using these
constants. Now we will go on to study the obtained solution under
different cosmological scenarios.

\subsection{Models for matter in the form of dust ($p_{m}=0$)}

The reconstruction of the dust fluid will be discussed in this
section so that the trace of energy momentum tensor
$\mathcal{T}=\rho_{m}$. Using equation (\ref{bb}) we get,
\begin{equation}
-T+f_{1}(T)-2 T f_{1T}=16 \pi G \mathcal{T}-f_{2}(\mathcal{T})+2
\mathcal{T} f_{2\mathcal{T}}=\lambda
\end{equation}
where the subscripts $T$ and $\mathcal{T}$ denote derivatives with
respect to $T$ and $\mathcal{T}$ respectively. The left and right
side completely depend on T and $\mathcal{T}$ respectively. As a
result of independence, both sides should be equal to a constant
(say $\lambda$) when using the separation of the variable method.
As a result, we get two ordinary equations,

\begin{equation}
-T+f_{1}(T)-2 T f_{1T}=\lambda
\end{equation}

\begin{equation}
16\pi G \mathcal{T}-f_{2}(\mathcal{T})+2 \mathcal{T}
f_{2\mathcal{T}}=\lambda
\end{equation}

The solutions of the above differential equations are obtained as,
\begin{equation}\label{flrw16}
f_{1}(T)=-T+\lambda+C_{1}\sqrt{T}
\end{equation}
and
\begin{equation}\label{flrw17}
f_{2}(\mathcal{T})=-\lambda-16 \pi G
\mathcal{T}+C_{2}\sqrt{\mathcal{T}}
\end{equation}
where $C_{1}$ and $C_{2}$ are constants of integration.
Substituting (\ref{flrw16}) and (\ref{flrw17}) in the functional
form of (\ref{flrw15}), we get
\begin{equation}\label{duss}
f(T,\mathcal{T})=-T+C_{1}\sqrt{T}-16\pi G
\mathcal{T}+C_{2}\sqrt{\mathcal{T}}
\end{equation}
Continuing from equation (\ref{fff}) we see that here the torsion
sector is unchanged. For the matter sector we have terms
proportional to $\mathcal{T}$ and $\sqrt{\mathcal{T}}$. Obviously
$\mathcal{T}$ is linear and not of much interest, but the term
involving $\sqrt{\mathcal{T}}$ can be of interest. Since here the
trace of the energy momentum tensor $\mathcal{T}$ is proportional
to the matter energy density $\rho_{m}$, so $\sqrt{\mathcal{T}}$
will be proportional to $\sqrt{\rho_{m}}$. We know that terms
involving $\sqrt{\rho_{m}}$ occur in the field equations of the
DGP brane model \cite{dvali} and loop quantum gravity \cite{lqg}.
Considering the utility and success of these theories the obtained
solution is quite interesting indeed. For $C_{2}=0$ this term
vanishes and we lose the probable flavours of the previously
mentioned gravity theories. Similarly for $C_{1}=0$ the Lagrangian
is a linear function in torsion.

\subsection{Models for matter in the form of perfect fluid with $p_{m}=-\frac{1}{3}\rho_{m}$}

In this case, we recreate the $f(T,\mathcal{T})$ Lagrangian for an
expanding universe (quintessence or dark energy era). Physically
intriguing is the value of the EoS parameter
$\omega=-\frac{1}{3}$, which is close to the upper limit of the
set of matter fields that obey the strong energy requirement. So
the trace of the energy momentum tensor is given by
$\mathcal{T}=2\rho_{m}$. Now from the equation (\ref{bb}), we get
\begin{equation}
-2T f_{1T}+f_{1}(T)-T=8\pi G
\mathcal{T}-f_{2}(\mathcal{T})+\frac{2}{3} \mathcal{T}
f_{2\mathcal{T}}=\mu
\end{equation}
where $\mu$ is a constant. Therefore, we get two differential
equations as
\begin{equation}
-2T f_{1T}+f_{1}(T)-T=\mu
\end{equation}
and
\begin{equation}\label{ae1}
8\pi G \mathcal{T}-f_{2}(\mathcal{T})+\frac{2}{3} \mathcal{T}
f_{2}(\mathcal{T})=\mu
\end{equation}

The solution of the above differential equations are obtained as,
\begin{equation}
f_{1}(T)=-T+\mu+C_{3}\sqrt{T}
\end{equation}

\begin{equation}
f_{2}(\mathcal{T})=-\mu+24 \pi G \mathcal{T}+C_{4}
\mathcal{T}^{\frac{3}{2}}
\end{equation}
where $C_{3}$ and $C_{4}$ are constants of integration.
Substituting $f_{1}(T)$ and $f_{2}(\mathcal{T})$ in
(\ref{flrw15}), we get the functional value as,
\begin{equation}
f(T,\mathcal{T})=-T+C_{3}\sqrt{T}+24 \pi G \mathcal{T}+C_{4}
\mathcal{T}^{\frac{3}{2}}
\end{equation}
The Lagrangian is similar to the one obtained in the previous case
eqn.(\ref{duss}) for the first three terms. The fourth term
containing the non-linear effects in $\mathcal{T}$ is somewhat
different. It has the term $\mathcal{T}\sqrt{\mathcal{T}}$ as
compared to the term $\sqrt{\mathcal{T}}$ in the previous case. So
we see that here the effect coming from the matter sector is more
pronounced compared to the dust case.

\subsection{Models for $\Lambda$CDM cosmology ($p_{m}=-\rho_{m}$)}

Here we consider the $\Lambda CDM$ cosmological scenario where the
equation of state parameter is given by $\omega=-1$. This serves
as a boundary for transition from a quintessence to a phantom
universe. Using equation (\ref{bb}) we get for this scenario,
\begin{equation}\label{mod1}
-T-2Tf_{1T}+f_{1}(T)=4\pi G \mathcal{T}-f_{2}(\mathcal{T})=\eta
\end{equation}
where $\eta$ is a constant. Using eqn.(\ref{mod1}) we get two
separate differential equations as,
\begin{equation}
-T-2Tf_{1T}+f_{1}(T)=\eta
\end{equation}
and
\begin{equation}
4\pi G \mathcal{T}-f_{2}(\mathcal{T})=\eta
\end{equation}

Solutions of the above differential equations are obtained as,
\begin{equation}
f_{1}(T)=-T+C_{5}\sqrt{T}+\eta
\end{equation}

\begin{equation}
f_{2}(\mathcal{T})=4\pi G \mathcal{T}-\eta
\end{equation}
where $C_{5}$ is the constant of integration. Substituting the
above expressions in the functional value (\ref{flrw15}), we get
\begin{equation}
f(T,\mathcal{T})=-T+C_{5}\sqrt{T}+4\pi G \mathcal{T}
\end{equation}
Here the Lagrangian is far more simpler compared to the previous
two cases. The torsional part is similar but only a linear term
occurs in $\mathcal{T}$. So it can be said that this solution is
almost a special case of the previous solutions by considering a
vanishing $C_{2}$ or $C_{4}$ and may be derivable from them.

\subsection{Einstein static universe in $f(T,\mathcal{T})$ gravity}

Here we have a vanishing Hubble parameter $H=0$, which gives
$T=-6H^{2}=0$. The necessary Friedmann equation (\ref{flrw9}) for
this scenario is obtained as,
\begin{equation}
16\pi G \rho_{m}-f+2 f_{\mathcal{T}}\left(\rho_{m}+p_{m}\right)=0
\end{equation}
If we take the matter content as dust $(p_{m}=0)$, this allows the
trace of energy momentum tensor to be $\mathcal{T}=\rho_{m}$,
simplifying the above equation to,
\begin{equation}
16\pi G \mathcal{T}-f(T,\mathcal{T})+2\mathcal{T}f_{\mathcal{T}}=0
\end{equation}
Considering the separable form of the functional form we separate
the variables as
\begin{equation}
16\pi
G\mathcal{T}+2\mathcal{T}f_{2\mathcal{T}}-f_{2}(\mathcal{T})=f_{1}(T)=\nu
\end{equation}

Solving the above equations we get,
\begin{equation}
f_{1}(T)=\nu
\end{equation}

\begin{equation}
f_{2}(\mathcal{T})=-16\pi G
\mathcal{T}-\nu+C_{6}\sqrt{\mathcal{T}}
\end{equation}
where $C_{6}$ is constant of integration. So the functional value
is obtained as,
\begin{equation}
f(T,\mathcal{T})=f(\mathcal{T})=-16\pi G
\mathcal{T}+C_{6}\sqrt{\mathcal{T}}
\end{equation}
This absence of the torsion term from the Lagrangian is quite
expected in this case. As the Hubble parameter vanishes for a
static universe, the torsion term vanishes as well and we are left
with a Lagrangian whose argument is limited to only the matter
sector, i.e. $\mathcal{T}$. This shows that in a static universe
the torsion component is rendered redundant or vice versa. As far
as the matter sector is concerned the solution is similar to the
solution obtained in the case of dust.

\section{Reconstruction of $f(T,\mathcal{T})$ Lagrangian for non-minimally matter coupled models}
Here we consider non-minimal coupling between the matter sector
and the torsion sector unlike the previous section where we
considered minimally coupled models. This scenario is more
realistic considering the fact that the whole idea of including
matter in the gravity Lagrangian is to allow for a non-minimal
matter coupling. These non-minimally coupled forms of the matter
derivative theories are conceptually different from the usual
general relativistic forms. These coupled forms are novel and
produce very interesting results quite different from the usual
results obtained in general relativity. We note that the trace of
the energy momentum tensor is given by
$\mathcal{T}=\rho_{m}\left(1-3\omega\right)$. Here we take into
consideration certain specific forms of the $f(T,\mathcal{T})$
functional as follows:- \vspace{5mm}

A)~ $T h(\mathcal{T})$~~~~~~ B)~ $\mathcal{T} g(T)$

 \vspace{5mm}

\subsection{$f(T,\mathcal{T})=T h(\mathcal{T})$}
The type of $f(T,\mathcal{T})$ that follows determines whether a
function $h(\mathcal{T})$ can satisfy a resizing of T. Here we
take $T$ and a function of $\mathcal{T}$ in a product form giving
a non-minimal coupling between the two components. Using
eqn.(\ref{flrw9}) we get the following equation for the scenario,
\begin{equation}
\frac{2(1+\omega)}{1-3\omega} T \mathcal{T} h_{\mathcal{T}}+T
h(\mathcal{T})+\frac{16\pi G}{1-3\omega}\mathcal{T}+T=0
\end{equation}
For convenience we consider a constant value of scalar torsion $T$
and obtain the solution of the above equation as,
\begin{equation}
h(\mathcal{T})=\frac{16\pi G
\mathcal{T}}{\left(\omega-3\right)T}-\left[1-{C_{7}\left\{2\mathcal{T}\left(1+\omega\right)\right\}}^\frac{-1+3\omega}{2\left(1+\omega\right)}\right]
\end{equation}
where $C_{7}$ is the constant of integration. So the functional
form is obtained as,
\begin{equation}\label{tedd}
f(T,\mathcal{T})=\frac{16\pi G \mathcal{T}
}{\omega-3}-T\left[1-{C_{7}\left\{2\mathcal{T}\left(1+\omega\right)\right\}}^\frac{-1+3\omega}{2\left(1+\omega\right)}\right]
\end{equation}
As expected this solution has a much richer structure than the
ones obtained for the separable cases in the previous section. We
see that we have a linear term in $T$, a linear term in
$\mathcal{T}$ and a coupled term containing a linear $T$ and a
non-linear $\mathcal{T}$. This last coupled term is of interest to
us because it allows for the non-minimal coupling between the
torsion and the matter sectors. Due to the presence of this term
it is expected that one can obtain a dark energy sector being
quintessence-like, phantom-like, or experiencing the
phantom-divide crossing during evolution just analogous to the
non-minimal teleparallel quintessence \cite{telde}.

\subsection{$f(T,\mathcal{T})=\mathcal{T} g(T)$}
We now examine the situation the trace of the energy momentum
tensor $\mathcal{T}$ is associated to any arbitrary function of
the torsion scalar $T$ in a product from. In this situation
eqn.(\ref{flrw9}) gives the required equation as
\begin{equation}\label{flrw19}
2T \mathcal{T}
g_{T}+\frac{\mathcal{T}(1+5\omega)}{1-3\omega}g(T)+T+\frac{16\pi G
\mathcal{T}}{1-3\omega}=0
\end{equation}
The solution of the above eqn.(\ref{flrw19}) is found as,
\begin{equation}
g(T)=-\frac{16\pi
G}{1+5\omega}+\frac{T\left(1-3\omega\right)}{\mathcal{T}\left(\omega-3\right)}+C_{8}\left[2T\left(1-3\omega\right)\right]^\frac{-1-5\omega}{2\left(1-3\omega\right)}
\end{equation}
where $C_{8}$ is the constant of integration. Using the above
solution the functional form of $f(T,\mathcal{T})$ can be given
by,
\begin{equation}
f(T,\mathcal{T})=-\frac{16\pi G
\mathcal{T}}{1+5\omega}+\frac{T\left(1-3\omega\right)}{\omega-3}+C_{8}\mathcal{T}\left[2T\left(1-3\omega\right)\right]^\frac{-1-5\omega}{2\left(1-3\omega\right)}
\end{equation}
Here again we have linear terms in $T$ and $\mathcal{T}$ and a
coupled term which has a linear $\mathcal{T}$ and a non-linear
$T$. Here again the presence of this non-minimally coupled term
may help us realize the phantom era and also the phantom crossing
as discussed in eqn.(\ref{tedd}). Here it should be noted that the
roles of $T$ and $\mathcal{T}$ will get reversed due to our choice
of the model as compared to the previous case.

Viability of these models will be based on the term
$\left[\mathcal{T}\left(1+\omega\right)\right]^{\frac{-1+3\omega}{2\left(1+\omega\right)}}$
for the model A and on the term
$\left[T\left(1-3\omega\right)\right]^{\frac{-1-5\omega}{2\left(1-3\omega\right)}}$
for the model B. Since similar terms appeared in the study of
section 5, we have presented the viability analysis in section
5.2. Model A will follow the same viability
analysis presented there. For model B we can see that there can be two cases.\\

$\bullet$ ~~$T\left(1-3\omega\right)>0$ implies that we can have
two possibilities.\\

$\star$~~~$T>0$ and $1-3\omega>0$ $\implies$ $\omega<1/3$. This
includes the region $\omega<-1/3$ which is compatible for studying
the accelerated expansion of the universe.\\

$\star$~~~$T<0$ and $1-3\omega<0$ $\implies$ $\omega>1/3$. This is
not useful in studying the late cosmic acceleration and is of no
use in the present context.\\

$\bullet$ ~~$T\left(1-3\omega\right)<0$ implies that we can have
two possibilities.\\

$\star$~~~$T>0$ and $1-3\omega<0$ $\implies$ $\omega>1/3$. This is
not useful in studying the late cosmic acceleration and is of no
use in the present context.\\

$\star$~~~$T<0$ and $1-3\omega>0$ $\implies$ $\omega<1/3$. This
includes the region $\omega<-1/3$ which is compatible for studying
the accelerated expansion of the universe. It should be kept in
mind that in this case we get real solution only when we take
integral values of the exponent
$\frac{-1-5\omega}{2\left(1-3\omega\right)}$. For non-integral
values we get imaginary values of the functional form which is
incompatible cosmologically.

\section{Cosmological solutions of $f(T,\mathcal{T})$ gravity}
The potential of obtaining gravitational Lagrangian
$f(T,\mathcal{T})$ that are suitable for modelling the expansion
of the universe as suggested by the power-law and de-sitter
solution is discussed in this section. We have also discussed a
third scenario where we have considered a hybrid model using the
power law and de-Sitter expansion models.

\subsection{Power law solutions}
It could be interesting to look into if the $f(T,\mathcal{T})$
gravity theory contains any specific energy solutions that
correlate to different stages of cosmic evolution. The scale
factor separates these solutions as the decelerated and
accelerated cosmic periods. We consider a power law form of scale
factor in terms of time $t$,
\begin{equation}\label{flrw20}
a(t)=a_{0}t^{m} , ~~~~   H(t)=\frac{m}{t},~ m>0
\end{equation}
where $a_{0}$ is a constant. The torsion scalar has the form~
$T=-6H^2=-6 m^{2} t^{-2}$~ for the matching scale factor.
Furthermore, the results of the conservation equation for
$p_{eff}=\omega_{eff} \rho_{eff}$ are
\begin{equation}\label{ed1}
\rho_{eff}(t)=\rho_{0} t^{-3m(1+\omega_{eff})}
\end{equation}
where $\rho_{0}$ is a constant. Using (\ref{flrw20}) and
(\ref{flrw15a}),we get
\begin{equation}\label{omef}
\omega_{eff}=\frac{2\dot{H}+3 H^2}{3H^2}=\frac{-2+3m}{3m}
\end{equation}
The effective density can be expressed in terms of the torsion
scalar as,
\begin{equation}
\rho_{eff}(T)=\rho_{0}\left(-\frac{6m^2}{T}\right)^{1-3m}
\end{equation}
We consider the functional form as
$f(T,\mathcal{T})=f_{1}(T)+f_{2}(\mathcal{T})$. Using the above
expressions in eqns.(\ref{flrw9}) and (\ref{flrw11}) we get,
\begin{equation}
8\pi G
\rho_{0}\left(-\frac{6m^2}{T}\right)^{1-3m}+\frac{f_{1}(T)}{2}-Tf_{1T}=8\pi
G
\rho_{m}-\frac{f_{2}(\mathcal{T})}{2}+f_{2\mathcal{T}}\left(\rho_{m}+p_{m}\right)=\xi
\end{equation}
where $\xi\neq 0$ is the separation constant. From the above
equation we obtain two separate differential equations,
\begin{equation}\label{diffo1}
8\pi G
\rho_{0}\left(-\frac{6m^2}{T}\right)^{1-3m}+\frac{f_{1}(T)}{2}-Tf_{1T}=\xi
\end{equation}
and
\begin{equation}\label{diffo2}
8\pi G
\rho_{m}-\frac{f_{2}(\mathcal{T})}{2}+f_{2\mathcal{T}}\left(\rho_{m}+p_{m}\right)=\xi
\end{equation}
Solving the first equation (\ref{diffo1}) we get,
\begin{equation}
f_{1}(T)=C_{9}\sqrt{T}+\frac{2^{5-3m}\times 27^{-m}G\pi \rho_{0}
}{(-m^2)^{3m-1}\left(2m-1\right)}T^{3m-1}+2\xi
\end{equation}
where $C_{9}$ is the constant of integration. Here we see that we
have two non-linear terms in $T$ and a constant term for the
geometric sector. The first term is proportional to $\sqrt{T}$
which is quite common in this study and appeared in almost all the
previous models. The second non-linear term is different in this
case and can be just anything depending on the value of $m$. So
here there is an increased degree of freedom in the geometric
sector compared to the previous models. The constant term can be
used to amplify the effect of the torsional sector where required.

The second differential equation (\ref{diffo2}) is not readily
solvable because it contains unknown quantities $\rho_{m}$ and
$p_{m}$. So we will consider certain cosmological scenarios to get
analytical solutions from this equation. The scenarios that need
to be
investigated are dust, perfect fluid and $\Lambda$CDM.\\

$\bullet$~~ Dust models ($p_{m}=0,~ \rho_{m}=\mathcal{T}$)

For this case eqn.(\ref{diffo2}) takes the form
\begin{equation}
8\pi G
\mathcal{T}-\frac{f_{2}(\mathcal{T})}{2}+\mathcal{T}f_{2\mathcal{T}}=\xi
\end{equation}
Solving the above equation we get,
\begin{equation}
f_{2}(\mathcal{T})=C_{10}\sqrt{\mathcal{T}}-16\pi G
\mathcal{T}-2\xi
\end{equation}
where $C_{10}$ is the constant of integration. Since energy
density of the universe is a positive value, so $\mathcal{T}$ must
also be positive. So the above solution yields real values and
hence may be compatible with the observations. Here the solution
obtained for the matter sector is quite similar to the ones
obtained for the previous cases except the constant term which may
be used to amplify the effect. Due to the first term the
DGP brane effect may be present here as discussed before.\\

$\bullet$~~ Perfect fluid models ($p_{m}=-1/3\rho_{m},~
\rho_{m}=\mathcal{T}/2$)

For this scenario eqn.(\ref{diffo2}) becomes,
\begin{equation}
4\pi G
\mathcal{T}-\frac{f_{2}(\mathcal{T})}{2}+\frac{\mathcal{T}}{3}f_{2\mathcal{T}}=\xi
\end{equation}
Solving the above equation we get,
\begin{equation}
f_{2}(\mathcal{T})=C_{11}\mathcal{T}^{3/2}+24\pi G
\mathcal{T}-2\xi
\end{equation}
where $C_{11}$ is the constant of integration. Here also we see
that the solution yields real values and hence this solution may
be observationally compatible. Here the solution is similar to the
dust case except the first term is amplified by a factor
$\mathcal{T}$. So here we can expect an amplified effect
of the matter sector.\\

$\bullet$~~ $\Lambda$CDM model ($p_{m}=-\rho_{m},~
\rho_{m}=\mathcal{T}/4$)

For this scenario eqn.(\ref{diffo2}) becomes,
\begin{equation}
2\pi G \mathcal{T}-\frac{f_{2}(\mathcal{T})}{2}=\xi
\end{equation}
solving which we get,
\begin{equation}
f_{2}(\mathcal{T})=4\pi G \mathcal{T}-2\xi
\end{equation}
This solution is also compatible with the observations. Finally we
have the simplest solution encountered in this work so far. This
is simply a linear term in $\mathcal{T}$ along with the expected
constant term. For a low density universe (due to expansion) a
linear term in matter density will dominate the non-linear terms.
In such a scenario this term will have significant effect on the
evolution of the universe. Finally we don't see
any DGP brane like effect in this case.\\

Similarly using using eqn.(\ref{flrw20}) in eqn.(\ref{decel}) we
get the deceleration parameter as,
\begin{equation}\label{decc}
q=\frac{1-m}{m}
\end{equation}
It is clearly seen that for $m>1$ we get the accelerated epoch,
whereas for $0<m<1$, there is a decelerated evolution of the
universe.

In Fig.(1) we have plotted the effective equation of state
parameter $\omega_{eff}$ against the parameter $m$ for the
power-law model. We see that the branch in the negative values of
$m$ is not physically viable since it gives positive values of
$\omega_{eff}$. For $0<m<0.5$ we see that the trajectory remains
in the negative level showing a dark energy dominated universe
($\omega_{eff}<-1/3$). This is observationally viable range of
$m$. In Fig.(2) we have generated the plot for the deceleration
parameter $q$ against the parameter $m$ for the power-law model.
We see that there is divergence around $m=0$. Other than that the
trajectory settles around $-1$ for both positive and negative
values of $m$. Negative value of $q$ indicates an accelerated
expansion of the universe. Moreover this shows that the model
asymptotically tries to mimic the de-Sitter like evolution.

From the Planck data we know that the current constrained value of
the deceleration parameter is -0.55 \cite{qqq}. We have plotted
the deceleration parameter in Fig.(2) against the model parameter
$m$. So it is difficult to understand the evolutionary epoch from
the plot. If it could have been plotted against the redshift
parameter $z$, it would have been easier to understand the current
epoch ($z=0$) and check the corresponding value of $q$ and compare
it with the current observed value. But this is not possible for
this plot. This plot just shows the trend and the negative value
of $q$ depicts an accelerated expansion. But it is obvious that we
can make suitable transformations to fit our result to the
observed value. From the plot we see that for $m=2$ we have
$q\approx -0.55$. So we can consider a suitable transformation to
match the current value of redshift $z=0$ with $m=2$ to make it
consistent with the observations.

\begin{figure}
~~~~~~~~~~~~~~~~~~~~~~~~~~~~~~\includegraphics[height=2.2in,width=3in]{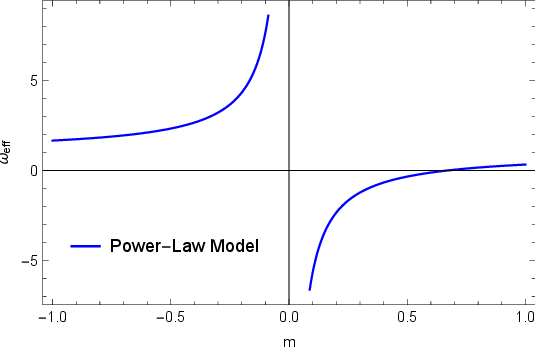}~~~~~\\

~~~~~~~~~~~~~~~~~~~~~~~~~~~~~~~~~~~~~~~~~~~~~~~~~~~~~~~~~~~~Fig.1~~~~~~~~~~~~~~~~~~~~\\

\vspace{1mm} \textit{\textbf{Fig.1} shows the variation of the
effective equation of state parameter $\omega_{eff}$ against the
parameter $m$ for the power-law model.}
\end{figure}

\begin{figure}
~~~~~~~~~~~~~~~~~~~~~~~~~~~~~~\includegraphics[height=2.2in,width=3in]{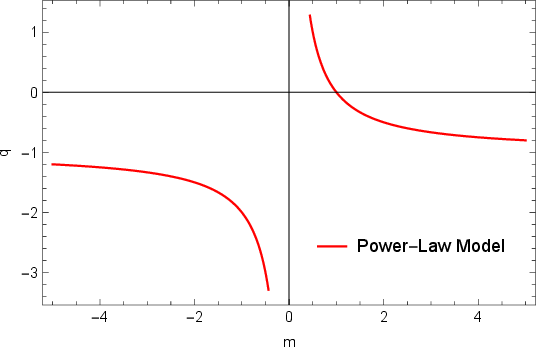}~~~~~\\

~~~~~~~~~~~~~~~~~~~~~~~~~~~~~~~~~~~~~~~~~~~~~~~~~~~~~~~~~~~~Fig.2~~~~~~~~~~~~~~~~~~~~\\

\vspace{1mm} \textit{\textbf{Fig.2} shows the variation of the
deceleration parameter $q$ against the parameter $m$ for the
power-law model.}
\end{figure}

\subsection{de-Sitter solutions}
A de-Sitter universe is a continuously expanding model of the
universe. Here the energy density of matter and radiation is
negligible as compared to that of dark energy. As a result this
model of universe is said to be an empty universe model devoid of
matter. The scale factor of this model increases exponentially and
the Hubble parameter takes a constant value. The scale factor is
taken as
\begin{equation}\label{desitacc}
a(t)=a_{0}e^{H_{0}t}, ~~~H(t)=H_{0}
\end{equation}
where $t$ is time and $a_{0}$ is a constant. In this model the
scale factor grows much faster than the power law model and in
future the universe is supposed to be dominated by the exponential
form of scale factor. For this model the scalar torsion is given
as $T=-6H_{0}^2$. We use the same functional form
$f(T,\mathcal{T})=f_{1}(T)+f_{2}(\mathcal{T})$ in
eqn.(\ref{flrw9}) and get the following differential equations,
\begin{equation}\label{desitt1}
f_{1}(T)+12H_{0}^{2}f_{1T}+6H_{0}^2=\kappa
\end{equation}
and
\begin{equation}\label{desitt2}
\frac{16\pi
G}{1-3\omega}\mathcal{T}-f_{2}(\mathcal{T})+2\left(\frac{1+\omega}{1-3\omega}\right)
\mathcal{T}f_{2\mathcal{T}}=\kappa
\end{equation}
Solving equation (\ref{desitt1}) we get,
\begin{equation}
f_{1}(T)=C_{12}e^{-\frac{T}{12H_{0}^2}}-6H_{0}^2+\kappa
\end{equation}
where $C_{12}$ is the constant of integration. Similarly solving
eqn.(\ref{desitt2}) we get,
\begin{equation}
f_{2}(\mathcal{T})=C_{13}\left[2\mathcal{T}\left(1+\omega\right)\right]
^{\frac{1-3\omega}{2\left(1+\omega\right)}}-\frac{16\pi
G}{1+5\omega}\mathcal{T}-\kappa
\end{equation}
where $C_{13}$ is the constant of integration. So the functional
becomes,
\begin{equation}
f(T,\mathcal{T})=C_{12}e^{-\frac{T}{12H_{0}^2}}+C_{13}\left[2\mathcal{T}\left(1+\omega\right)\right]
^{\frac{1-3\omega}{2\left(1+\omega\right)}}-\frac{16\pi
G}{1+5\omega}\mathcal{T}-6H_{0}^2
\end{equation}
This solution has many interesting features. The first term is the
geometric term which is exponential in $T$. So highly non-linear
and is responsible for a de-Sitter like expansion. The second and
the third terms are respectively non-linear and linear in
$\mathcal{T}$ which have their usual effect as discussed before in
the previous models. These are obviously dependent on the value of
$\omega$ for their ultimate effect. The final constant term is
interpreted as the present value of the Hubble parameter which
changes with the evolution of the universe. So the current issue
of Hubble tension will have a lot of effect on this solution due
to this last term.\\

$\bullet$~~ Viability of the solution\\

There can be two cases of discussion. We can have
$\mathcal{T}\left(1+\omega\right)>0$ or
$\mathcal{T}\left(1+\omega\right)<0$.\\

$\star$~~  $\mathcal{T}\left(1+\omega\right)>0$ implies two
sub-cases.\\

In the first subcase we have
$\mathcal{T}=\left(1-3\omega\right)\rho_{m}>0$ and $1+\omega>0$.
This gives $\omega<1/3$ and $\omega>-1$. The intersection of the
two gives $-1<\omega<1/3$. The sub-region $-1<\omega<-1/3$ is
viable for studying an accelerated universe, however the
sub-region $-1/3<\omega<1/3$ is not quite viable.\\

In the second sub-case we have
$\mathcal{T}=\left(1-3\omega\right)\rho_{m}<0$ and $1+\omega<0$.
This gives $\omega>1/3$ and $\omega<-1$. This is not possible
simultaneously and so no viable scenario is obtained.\\

$\star$~~  $\mathcal{T}\left(1+\omega\right)<0$ implies two
sub-cases.\\

In the first subcase we have
$\mathcal{T}=\left(1-3\omega\right)\rho_{m}>0$ and $1+\omega<0$.
This gives $\omega<1/3$ and $\omega<-1$. The intersection of the
two gives $\omega<-1$. This is perfectly viable for studying an
accelerated universe. However this scenario can only be studied
for integral values of $\frac{1-3\omega}{2\left(1+\omega\right)}$.
This is because for non-integral values we do not get real values
of the functional form.\\

In the second sub-case we have
$\mathcal{T}=\left(1-3\omega\right)\rho_{m}<0$ and $1+\omega>0$.
This gives $\omega>1/3$ and $\omega>-1$. The intersection of the
two gives $\omega>1/3$. This is not useful for studying an
accelerated scenario.\\

For the de-Sitter model we see that the Hubble parameter takes a
constant value. So its derivative will vanish giving a constant
value to the deceleration parameter. Using eqn.(\ref{desitacc}) in
eqn.(\ref{decel}) we get the deceleration parameter for this model
as $q=-1$. This gives a constantly accelerating universe.

\subsection{Hybrid model}
Here we consider a hybrid model using the above power law model
and the exponential model. A hybrid model will help in probing the
combined effects of the above discussed models and can also
present new interesting solutions which may be more compatible
with the observations. We consider a product form of coupling
between the two models and get the scale factor as \cite{hybmod1,
hybmod2},
\begin{equation}\label{hybacc}
a(t)=a_{0}t^{n}e^{mt},~~~H(t)=m+\frac{n}{t}, ~~n\geq0
\end{equation}
where $t$ is time and $a_{0}$ is a constant. We see that for
$n=0$, we get the de-Sitter universe and for $m=0$ we get the
power law form of expansion. It is obvious from eqn.(\ref{hybacc})
that, the power law behaviour dominate the cosmic dynamics in
early phase of cosmic evolution and the exponential factor
dominates at the later phase. In \cite{hybmod1} the authors have
shown that the hybrid scale factor fosters an early deceleration
as well as a late-time acceleration and mimics the present
universe. In \cite{hybmod2} the authors found that the hybrid
scale factor simulates a cosmic transit behaviour from a
decelerated phase of expansion to an accelerated phase. The
torsion scalar is determined as $T=-6(m+\frac{n}{t})^{2}$.
Furthermore, the results of the conservation equation for
$p_{eff}=\omega_{eff} \rho_{eff}$ are
\begin{equation}\label{ed1ll}
\rho_{eff}(t)=\rho_{1}
e^{-3\left(1+\omega_{eff}\right)\left(mt+n\log t\right)}
\end{equation}
where $\rho_{1}$ is a constant. The effective EoS parameter is
given by,
\begin{equation}
\omega_{eff}=\frac{2\dot{H}+3 H^2}{3H^2}=1-\frac{2n}{3(mt+n)^{2}}
\end{equation}
We find the $\rho_{eff}$ in terms of scalar torsion as,
\begin{equation}
\rho_{eff}(T)=\rho_{1}e^{-3n\left[2-\frac{2\left(m\sqrt{-6T}+T\right)}{3nT}\right]\left[\frac{6m}{\sqrt{-6T}-6m}+\log\left(\frac{6n}{\sqrt{-6T}-6m}\right)\right]}
\end{equation}

Considering the functional form
$f(T,\mathcal{T})=f_{1}(T)+f_{2}(\mathcal{T})$ we proceed to find
solutions for this model. Using the above expressions in
eqns.(\ref{flrw9}) and (\ref{flrw11}) we get two differential
equations as,
\begin{equation}\label{hyb11}
8\pi
G\rho_{1}e^{-3n\left[2-\frac{2\left(m\sqrt{-6T}+T\right)}{3nT}\right]\left[\frac{6m}{\sqrt{-6T}-6m}+\log\left(\frac{6n}{\sqrt{-6T}-6m}\right)\right]}+\frac{f_{1}(T)}{2}-Tf_{1T}=\zeta
\end{equation}
and
\begin{equation}\label{hyb12}
8\pi
G\rho_{m}-\frac{f_{2}(\mathcal{T})}{2}+f_{2\mathcal{T}}\left(\rho_{m}+p_{m}\right)=\zeta
\end{equation}
where $\zeta$ is the separation constant. We do not get a direct
analytic solution from eqn.(\ref{hyb11}). So we expand the
exponential term in series and consider the linear terms only.
With this simplification we get the solution as,

$$f_{1}(T)=\frac{1}{3m\sqrt{-T}}\times\left[3m\left\{-16\sqrt{6}Gm\pi
\rho_{1}+2\sqrt{-T}\left(\zeta-48Gn\pi
\rho_{1}\right)+C_{14}\sqrt{-T}\right\}\right.$$
\begin{equation}
\left.+4\pi
G\rho_{1}\left\{12m\left(\sqrt{6}m+2\left(3n-1\right)\sqrt{-T}\right)
\log\left(\frac{6n}{-6m+\sqrt{-6T}}\right)
+\sqrt{6}T\left\{2\log\left(-6m+\sqrt{-6T}\right)-\log(-T)\right\}\right\}\right]
\end{equation}
where $C_{14}$ is the constant of integration. Equation
(\ref{hyb12}) is similar to eqn.(\ref{diffo2}) obtained in the
case of power law models. So we will use the solutions
obtained in that section for different cosmological scenarios.\\\\

$\bullet$~~ For dust models ($p_{m}=0,~~\rho_{m}=\mathcal{T}$) we
get the functional form as:

$$f(T,\mathcal{T})=\frac{1}{3m\sqrt{-T}}\times\left[3m\left\{-16\sqrt{6}Gm\pi
\rho_{1}+2\sqrt{-T}\left(\zeta-48Gn\pi
\rho_{1}\right)+C_{14}\sqrt{-T}\right\}\right.$$

$$\left.+4\pi
G\rho_{1}\left\{12m\left(\sqrt{6}m+2\left(3n-1\right)\sqrt{-T}\right)
\log\left(\frac{6n}{-6m+\sqrt{-6T}}\right)
+\sqrt{6}T\left\{2\log\left(-6m+\sqrt{-6T}\right)-\log(-T)\right\}\right\}\right]$$

\begin{equation}
+C_{15}\sqrt{\mathcal{T}}-16\pi G \mathcal{T}-2\zeta
\end{equation}
where $C_{14}$ and $C_{15}$ are constants of integration. This
solution is different considering that it contains terms with
$\sqrt{-T}$ which we did not encounter previously. The first and
the second terms are coming from the geometric sector and are
quite lengthy. The second term even contains various logarithmic
terms. Expanding the logarithmic terms in series we can have a
simplified form of this part but we will need to truncate the
series which will bring in approximation. The last three terms
arise from the matter sector and have been dealt previously.
\\\\

$\bullet$~~ For Perfect fluid models ($p_{m}=-1/3\rho_{m},~
\rho_{m}=\mathcal{T}/2$) we get the functional form as:

$$f(T,\mathcal{T})=\frac{1}{3m\sqrt{-T}}\times\left[3m\left\{-16\sqrt{6}Gm\pi
\rho_{1}+2\sqrt{-T}\left(\zeta-48Gn\pi
\rho_{1}\right)+C_{14}\sqrt{-T}\right\}\right.$$

$$\left.+4\pi
G\rho_{1}\left\{12m\left(\sqrt{6}m+2\left(3n-1\right)\sqrt{-T}\right)
\log\left(\frac{6n}{-6m+\sqrt{-6T}}\right)
+\sqrt{6}T\left\{2\log\left(-6m+\sqrt{-6T}\right)-\log(-T)\right\}\right\}\right]$$

\begin{equation}
+C_{16}\mathcal{T}^{3/2}+24\pi G \mathcal{T}-2\zeta
\end{equation}
where $C_{14}$ and $C_{16}$ are constants of integration.
\\\\

$\bullet$~~ For the $\Lambda$CDM model ($p_{m}=-\rho_{m},~
\rho_{m}=\mathcal{T}/4$) the functional form is obtained as:

$$f(T,\mathcal{T})=\frac{1}{3m\sqrt{-T}}\times\left[3m\left\{-16\sqrt{6}Gm\pi
\rho_{1}+2\sqrt{-T}\left(\zeta-48Gn\pi
\rho_{1}\right)+C_{14}\sqrt{-T}\right\}\right.$$

$$\left.+4\pi
G\rho_{1}\left\{12m\left(\sqrt{6}m+2\left(3n-1\right)\sqrt{-T}\right)
\log\left(\frac{6n}{-6m+\sqrt{-6T}}\right)
+\sqrt{6}T\left\{2\log\left(-6m+\sqrt{-6T}\right)-\log(-T)\right\}\right\}\right]$$

\begin{equation}
+4\pi G \mathcal{T}-2\zeta
\end{equation}
\\
For both the above cases the solutions are similar to the dust
case in the geometric sector. Subtle changes appear in the matter
terms for the different cosmological scenario but they have
already been discussed before. The important thing to note in
these solutions is the complexity of the geometric sector which
need a lot of approximation before we can get any clear idea.
Since $T=-6H^2$, $\sqrt{-6T}$ and $\log(-T)$ in the above
expressions will be real and these functional forms can be
mathematically and observationally viable.

Using eqn.(\ref{hybacc}) in eqn.(\ref{decel}) we get the
deceleration parameter for this model as,
\begin{equation}\label{hybdec}
q=-1+\frac{n}{\left(mt+n\right)^2}
\end{equation}
Moreover in terms of the torsion scalar the deceleration parameter
can be expressed as,
\begin{equation}
q=-1+\frac{1}{n\left(\frac{m}{\sqrt{-T/6}-m}+1\right)^{2}}
\end{equation}
To realize an accelerating scenario we should have
$T>-\frac{6m^2}{\left(1-\sqrt{n}\right)^{2}}$. From
eqn.(\ref{hybdec}) we see that at an early phase of cosmic
evolution, the deceleration parameter becomes $-1+\frac{1}{n}$ and
and at the later phase it approaches $-1$. This shows that the
universe goes from a power law phase to a de-Sitter phase. This
fact is demonstrated in Fig.(4).

In Fig.(3), we have generated a plot for the effective equation of
state parameter $\omega_{eff}$ against time $t$ for different
values of the parameters $m$ and $n$ for the hybrid model. It is
seen that with the evolution of time there is a transition of the
trajectory from the positive to the negative level. This clearly
shows that the universe undergoes a transition from the matter
dominated to the dark energy dominated era. This is compatible
with the late cosmic acceleration. Similarly in Fig.(4) we have
plotted the deceleration parameter $q$ against time $t$ for
different values of the parameters $m$ and $n$ for the hybrid
model. It is seen that as the universe evolves $q$ undergoes a
transition from positive to negative level. This shows that the
late time universe is undergoing an accelerated expansion. This
signature flipping behaviour of the deceleration parameter is an
evolutionary aspect of this model as given in \cite{hybmod2}. This
is completely compatible with the observations. Here we see that
around $t\approx 0.2$ we have $q=-0.55$. So for some suitable
transformation $t\approx 0.2$ can represent the current universe
($z=0$) and our result matches with the observations. Similarly
the timing of signature flipping i.e. the transition from an
decelerated to an accelerated universe can be matched with the
observed redshift value $z=0.45$.

\begin{figure}
~~~~~~~~~~~~~~~~~~~~~~~~~~~~~~\includegraphics[height=2.2in,width=3in]{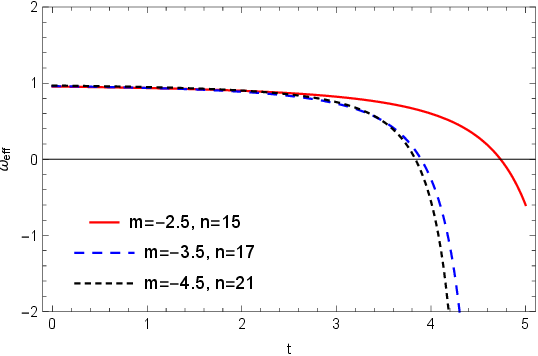}~~~~~\\

~~~~~~~~~~~~~~~~~~~~~~~~~~~~~~~~~~~~~~~~~~~~~~~~~~~~~~~~~~~~Fig.3~~~~~~~~~~~~~~~~~~~~\\

\vspace{1mm} \textit{\textbf{Fig.3} shows the variation of the
effective equation of state parameter $\omega_{eff}$ against time
$t$ for different values of the parameters $m$ and $n$ for the
hybrid model.}
\end{figure}

\begin{figure}
~~~~~~~~~~~~~~~~~~~~~~~~~~~~~~\includegraphics[height=2.2in,width=3in]{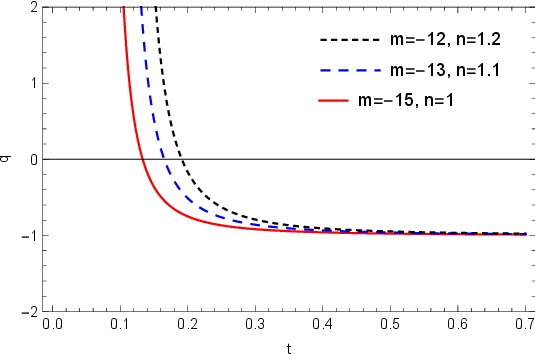}~~~~~\\

~~~~~~~~~~~~~~~~~~~~~~~~~~~~~~~~~~~~~~~~~~~~~~~~~~~~~~~~~~~~Fig.4~~~~~~~~~~~~~~~~~~~~\\

\vspace{1mm} \textit{\textbf{Fig.4} shows the variation of the
deceleration parameter $q$ against time $t$ for different values
of the parameters $m$ and $n$ for the hybrid model.}
\end{figure}

%\subsubsection{Model-II}
%\begin{equation}
%a(t)=a_{1}t^{n}+a_{2}e^{mt},~
%H(t)=\frac{a_{1}nt^{n-1}+a_{2}me^{mt}}{a_{1}t^{n}+a_{2}e^{mt}},~~m,n>0
%\end{equation}
%The torsion scalar becomes
%$T=-6\left(\frac{a_{1}nt^{n-1}+a_{2}me^{mt}}{a_{1}t^{n}+a_{2}e^{mt}}\right)^{2}$.
%The effective EoS parameter is given by,
%\begin{equation} \omega_{eff}=\frac{2\dot{H}+3
%H^2}{3H^2}=\frac{a_{1}^{2}n\left(3n-2\right)t^{2n}+3a_{2}^{2}m^{2}t^{2}e^{2mt}
%+2a_{1}a_{2}t^{n}e^{mt}\left\{n\left(n-1\right)+mnt+m^{2}t^{2}\right\}}
%{3\left(a_{1}nt^{n}+a_{2}mte^{mt}\right)^{2}}
%\end{equation}

\section{Conclusion}
In this work we have presented a reconstruction mechanism in the
background of $f(T,\mathcal{T})$ gravity theory for different
cosmological scenarios as well as for different forms of the
Lagrangian functional. The novel features of the modified gravity
theory based on scalar torsion and the energy momentum tensor
trace are discussed in detail. The field equations are revisited
and also important cosmological parameters like the equation of
state parameter and deceleration parameter are discussed in the
background of the $f(T,\mathcal{T})$ gravity theory.

A separable form of the Lagrangian functional in terms of the
torsion scalar and the energy momentum tensor trace is considered
for simplicity, but without any loss of generalization.
Considering different cosmological scenarios such as dust, perfect
fluid, $\Lambda$CDM, and Einstein static universe we have
calculated the reconstructed Lagrangian functional of the
$f(T,\mathcal{T})$ gravity theory both for the minimally and
non-minimally matter coupled models. Different structural forms
(both separable and coupled) of the functional were taken as a
priori to facilitate the mathematical process. Viability of these
models are also discussed to check whether they will be
mathematically real when combined with observational data. This
part basically helps us to explore the role of the equation of
state parameter in the reconstructed Lagrangian.

In section 5 we have studied some important cosmological solutions
in $f(T,\mathcal{T})$ gravity theory. Different forms of the
cosmological scale factors are considered such as the power law
model, de-Sitter model and a combination of these two. Using these
forms, $f(T,\mathcal{T})$ Lagrangian is reconstructed for a
separable form of priori model. Different cosmological scenarios
such as dust, perfect fluid, $\Lambda$CDM are considered wherever
required for simplification. Viability of the solutions are
discussed for all possible scenarios. Plots have been generated
for important cosmological parameters such as the equation of
state parameter and the deceleration parameter to check the
compatibility of the models with the observations. It is found
that the models support the late cosmic acceleration and are
cosmologically viable with the observations. The Lagrangian
functions obtained in section 3, 4 and 5 were discussed in detail
in the context of both mathematical and physical implications. It
was seen that in many models DGP brane like terms appeared which
is an interesting thing coming from this study. The square root
terms of the torsion scalar are basically the DGP terms. This is
because the terms evolve like the DGP gravity. We note that the
square root terms of torsion scalar are in fact boundary terms and
may be removed \cite{bound1, bound2}. This is an important
property of the solutions found. These terms basically creep up
from the integration by parts and sometimes vanish on the
application of boundary conditions. Nevertheless it is understood
that these square root terms should have minimal effect on the
solution even if they are present. It can be seen that in most of
the solutions apart from the these boundary terms we have another
linear term in torsion scalar. If we remove the square root term
then the expression actually becomes linear in the torsion scalar.
This is an important thing to note regarding the solutions. Since
our functional forms are not very complicated, we think it will
not be bad to keep these boundary terms in the expressions.
However it should be remembered that since they are boundary terms
they can be eliminated as per requirement any time and these terms
will have minimal effect on the qualitative and quantitative
features of the models.

The Lagrangian functional forms calculated in this work can
further be used in other studies to check their cosmological
compatibility. An immediate project that can be undertaken is to
constrain these Lagrangians using observational data like the
Hubble data, Planck data, Bicep data, BAO, CMB, etc. This work can
be a good sequel of the present work and will be attempted in a
following project. Another important work will be to explore a
perturbation analysis using the constructed Lagrangians to check
the stability of the models. One can also perform a dynamical
system analysis and check whether the models posses a rich phase
space structure.

%%%%%%%%%%%%%%%%%%%%%%%%%%
\section*{Acknowledgments}
%%%%%%%%%%%%%%%%%%%%%%%%%%

P.R. and F.R. acknowledges the Inter University Centre for
Astronomy and Astrophysics (IUCAA), Pune, India for granting
visiting associateship. We thank the anonymous referee for his/her
invaluable comments which helped us to improve the quality of the
manuscript.

%%%%%%%%%%%%%%%%%%%%%%%%%%%%%%%%%%%%%%%%%%%%
\section*{Data Availability Statement}
%%%%%%%%%%%%%%%%%%%%%%%%%%%%%%%%%%%%%%%%%%%%

No new data were generated or analyzed in this paper.

%%%%%%%%%%%%%%%%%%%%%%%%%%%%%%%%%%%%
\section*{Conflict of Interest}
%%%%%%%%%%%%%%%%%%%%%%%%%%%%%%%%%%%%

There are no conflicts of interests in this paper.

\end{document}